%% file: Monjo_2017_inhomogeneous_cosmology_v4.tex
\documentclass[aps,prl,twocolumn,showpacs,superscriptaddress,groupedaddress]{revtex4}  
\usepackage{graphicx}  
\usepackage{dcolumn}   
\usepackage{bm}        
\usepackage{amssymb}   
\usepackage{standalone}
\usepackage{filehook}
\usepackage{gincltex}
\usepackage{collectbox}
\usepackage{filemod-expmin}
\usepackage{cancel}
\usepackage{amsmath}
\usepackage{tikz}

\begin{document}


\widetext

\pagenumbering{arabic}



\title{Study of the observational compatibility of an inhomogeneous cosmology with linear expansion according to SNe Ia}
\input author_list.tex       
\date{\today}

\begin{abstract}
Most of current cosmological theories are built combining an isotropic and homogeneous manifold with a scale factor that depends on time. If one supposes a hyperconical universe with linear expansion, an inhomogeneous metric can be obtained by an appropriate transformation that preserves the proper time. This model locally tends to a flat Friedman-Robertson-Walker metric with linear expansion. The objective of this work is to analyse the observational compatibility of the inhomogeneous metric considered. For this purpose, the corresponding luminosity distance was obtained and was compared with the observations of 580 SNe Ia, taken from the Supernova Cosmology Project (SCP). The best fit of the hyperconical model obtains $\chi_0^2 = 562$, the same value that the standard $\Lambda$CDM model. Finally, a possible relationship is found between both theories. 
\end{abstract}

\pacs{98.80.Es, 98.80.Jk}

\maketitle

\section{I. Introduction}
\label{intro}
\subsection{A. Motivation}
\label{sec:Motivation}
\citet{RefJ} used the analogy of a balloon to explain Hubble's law. According to this heuristic model, if we draw galaxies on a balloon surface that is inflating, the galaxies are separating from each other in a similar way as our universe is expanding. Therefore, parallelism is carried out between the balloon 2-surface (2D) and the 3-surface (3D) of our universe, with the radius expanding as a function of time. The balloon expansion implies a time dimension, and its curved 3-surface implies that it is contained in a larger dimensional space. The balloon model of Eddington corresponds to a 3-spherical universe embedded in a 4+1 space-time manifold.

Modern cosmology is based on General Relativity (GR), i.e. it is constructed on pseudo-Riemannian manifolds $M$ with a Lorentzian metric \textit{g}, usually noted as ($M$,$g$). For instance, let $M \subseteq \mathbb{R}^{n+1}$ be an $(n+1)$-dimensional Lorentzian manifold with $n$ spatial dimensions and one temporal. This work uses the $(1, n)$ signature referring to one positive and $n$ negative eigenvalues of the metric, and notes $\mathbb{R}_{\eta}^{1,n} := (\mathbb{R}^{n+1}, \eta_{1,n})$ for the Minkowskian manifold (typically $n=3$) with this signature, i.e. with a flat diagonal metric $\eta_{1,n} := diag(1,-1,...,-1)$. 

Most of observations support the standard cosmology, including the Cosmic Microwave Background (CMB) and type Ia Supernovae (SNe Ia). However it requires some assumptions to explain the known ``horizon and flatness problems'' and why the universe began to expand. Particularly, \textit{inflation theory} attempts to complete the cosmology theory using a hypothetical super-rapid expansion in the early stages of the universe, which could favour the high homogeneity and isotropy of the universe today. However, this theory presents some problems: According to \citet{RefB, RefC}, an inflationary universe is much less probable than a non-inflationary universe, and \citet{A1, A2} argue that no observations support the validity of this theory.

Current cosmological theories are based on the $\Lambda$CDM model, constructed using the Friedmann equations and the Friedmann-Lema{\^\i}tre-Robertson-Walker (FRW) metric. This model requires fitting several parameters according to cosmological observations \citep{A3}. The most important fitted parameter is the cosmological constant, which is related to \textit{dark energy}.

Modified Gravity (MOG) theories are being tested in order to find alternatives to dark energy, but they are still inconclusive \citep{Clifton1, Koyama1}. Moreover, they are resurging cosmological theories based on fewer parameters, such as the linearly expanding model, which seem to be compatible with observations \citep{Milne1, Melia1}. In this sense, Eddington's idea about the embedded universe can be useful to develop these unconventional cosmologies.

\subsection{B. Problems of the current theory}
\label{sec:Problems}
Despite the good agreement with CMB measurements from international projects such as COBE, WMAP and Planck, some problems are found in the current cosmological theories. Probably, the most important problem of the $\Lambda$CDM theory is the no-direct detection of the cosmological constant, or its related dark energy, outside the framework of the theory. Despite the high precision of recent cosmological measurements, some authors describe tensions between several measurements when interpreted within the $\Lambda$CDM models \citep{A4, A5}. In fact, important errors could occur when applying General Relativity to cosmology, as the formulation of this theory does not guarantee that it is valid at cosmological scale. Another limitation of the Friedmann Equations is that they assume the universe can be approximated as a perfect fluid, but this is not the case in general. 

In addition, the Friedmann-Lema{\^\i}tre-Robertson-Walker (FRW) universes consider spatial and temporal factors are separable. In particular, spatial coordinates $\vec{\ell}$ can be written using comoving coordinates $\vec{\ell}^{\prime}$ and a scale factor $a(t)$ varying with time $t$. However, some problems are found for embedded manifolds. For example, if the universe is characterised by an expanding $3$-sphere $S_R^{3} := \left\{ \vec{\ell} \in  \mathbb{R}^{4} : |\vec{\ell}| = R(t) \right\}$ of radius $R(t)$ and centred in the origin (singularity in $t = 0$), the scale factor is $a(t) = {R(t)}/{R(t_{o})}$ and total spatial coordinates are $\vec{\ell} = a(t){\vec{\ell}}^{\prime} \in S_R^{3}$. Therefore, the line element contains both spatial and temporal differentials, i.e., $d\ell = \textit{a}(\textit{t})d\ell^{\prime} + \ell^{\prime} da(t)$. However, the FRW universes assume the spatial differential does not contain any temporal differential: $d\vec{\ell} = a(t)d\vec{\ell}^{\prime}$. In this way, the FRW metrics have differential lines: 
\begin{eqnarray} 
ds^2 = dt^2 - d\vec{\ell}^2 = dt^2 - a(t)^{2}{d}{{\vec{\ell}}^{\prime}}{}^{2} 
\end{eqnarray}   
Using an adequate transformation, the additional temporal term $\ell^{\prime} da(t)$ can be absorbed into the spatial term $\textit{a}(\textit{t})d\ell^{\prime}$, but producing a radial inhomogeneity (see Sect. II). Therefore, even for the simple case of a 3-sphere expanding as $R(t) = t$, the (isotropic and homogeneous) embedded manifolds cannot be described by the FRW metric due to the resulting inhomogeneity (under the intrinsic view).

On the other hand, several coincidences cannot be explained by the standard model: Why is the temperature isotropic in regions initially disconnected (``horizon problem'')? Why is energy density approximately equal to the critical density (``flatness problem'')? Why is the age of the universe close to the Hubble time? What is the origin of dark energy? Is it possible that other models can explain all these issues?

The apparent coincidence between the current age of the universe ($t_0$) and the current size of the Hubble sphere ($1/H_0$) has been recognised by other authors \citep{Melia1, Lima1, Melia2}. For this reason, alternative cosmological models are based on the equality $1/H = t$, such as the Milne or Dirac-Milne model or Melia's universe \citep{Milne1, Melia3}. These models have been criticised because they predict an equation of state of $w = -1/3$, however the criticisms are based on the assumption that Friedmann equations are valid (and they are not necessarily valid) \citep{Lewis1}. For instance, \citet{Jimenez2009} demonstrated that observations of the universe's flatness are based on the assumption of validity of the current model and, changing the frame theory, the same observations can be explained using another geometry, even with positive curvature. 

This work focuses on the same simple hypothesis (linear expansion), but analyses the curvature tensor according to the view of a hypothetical observer located in the hypersurface of an expanding universe of positive or negative curvature. This contrasts with the Dirac-Milne universe, which assumes a negative curvature in a classic FRW metric \citep{Milne1}. The starting point of the new metric is the construction of a 4-dimensional Lorentzian manifold embedded into the flat 5-dimensional Minkowskian space-time.

The aim of this work is twofold. Firstly, a new method is presented for obtaining radial inhomogeneous metrics by using the extrinsic view of an (embedded) homogeneous and isotropic universe with linear expansion. The inhomogeneity is built through an appropriate transformation that preserves the \textit{proper time}. Secondly, the model is tested comparing the theoretical and empirical luminosity distance using data from Type Ia supernovae.

\subsection{C. $\Lambda$CDM model}
\label{sec:LCDM}
A reminder of the standard cosmological theory is required in this work in order to present a comparison with the new results. FRW metric is obtained by combining a manifold with constant spatial curvature $K$ and a scale factor $a(t)$, which expands with time $t$. One can suppose the spatial section of the universe is a 3-sphere $S^{3}$ of curvature $K = {1}/{R^{2}}$, or a 3-hyperboloid H${}^{3}$ of curvature $K = -{1}/{R^{2}}$, where $R$ is the radius of curvature (i.e., a general curvature is $K = \pm{1}/{R^{2}}$). The Euclidean space (flat universe) can be taken as the limit as $K$ approaches zero. With this, the square of the line element in comoving spherical coordinates ($t$, $r^{\prime}$, $\theta$ ,$\phi$) is:
\begin{eqnarray} \label{eq:difFRW}   
ds^2 = dt^2 - a(t)^2\left(\frac{d{r'}^2}{1 - K{r'}^2} + {{r'}^{2}}d{\Sigma}^2\right)
\end{eqnarray} 
where $d{\it{\Sigma}}^2 := d\theta^2 + sin^2\theta d\phi^2$. Scale factor $a(t)$ represents the expansion of the universe, and hence the Hubble parameter is defined as $H := \dot{a}/a$. The Ricci tensor $R_{\alpha \beta}$ is calculated using the metric $g_{\alpha \beta}$ deduced from Eq.~\ref{eq:difFRW}, and thus, Einstein field equations are obtained according to:
\begin{eqnarray} \label{eq:RicciFRW}   
R_{\alpha \beta} - \frac{1}{2}g_{\alpha \beta}R = 8{\pi}\mathrm{G}T_{\alpha \beta} + \Lambda g_{\alpha \beta}
\end{eqnarray} where $R$ is the Ricci curvature scalar, G is the gravitational constant, $\Lambda$ is the cosmological constant and $T_{\alpha \beta}$ is the stress-energy tensor. If one assumes an homogeneous perfect fluid at rest (of density $\rho$ and pressure $p$) and calculates the Ricci tensor, the following Friedmann equations are obtained: 
\begin{eqnarray} \label{eq:Friedmann.eq1}   
\left(\frac{\dot{a}}{a} \right)^2 = \frac{8{\pi}{\mathrm{G}}\rho}{3} + \frac{\Lambda}{3} - \frac{K}{a^2} 
\end{eqnarray} 
\begin{eqnarray} \label{eq:Friedmann.eq2}   
2\left(\frac{\ddot{a}}{a} \right) +\left(\frac{\dot{a}}{a} \right)^2 + \frac{K}{a^2} - {\Lambda} = -{8{\pi}{\mathrm{G}}p} 
\end{eqnarray}

If critical density is defined as $\rho_{crit} := {3{H_{0}}^{2}}/{8\pi G}$, where $H_{0}$ is the current value for the Hubble parameter, Eq.~\ref{eq:Friedmann.eq1} is usually rewritten as:
\begin{eqnarray} \label{eq:Friedmann.eq1b}   
\frac{{H}^2}{{{H}_{0}}^2}  = \frac{\rho}{{\rho}_{crit}} + \frac{{\rho}_{\Lambda}}{\rho_{crit}} - \frac{K}{{{H}_{0}}^2 a^2} 
\end{eqnarray} where ${\rho}_{\Lambda}$ is the dark energy density, defined as ${\rho}_{\Lambda} := {\Lambda}/{8\pi\mathrm{G}}$. Assuming that matter density $\rho$ is cold (non ultra-relativistic), its equation of state $p = w\rho$ is approximately $w \approx 0$, and it varies by the expansion as $a^{-3}$. If the universe is dominated by radiation or ultra-relativistic particles, $w \approx 1/3$ and $\rho$ varies as $a^{-4}$.  Therefore, Eq.~\ref{eq:Friedmann.eq1b} can be rewritten as:
\begin{eqnarray} \label{eq:Friedmann.eq1c}   
\frac{{H}^2}{{{H}_{0}}^2}  = {\Omega}_{r} \left(\frac{{a}_{o}}{a} \right)^{4} + {\Omega}_{m} \left(\frac{{a}_{o}}{a} \right)^{3} + {\Omega}_{K} \left(\frac{{a}_{o}}{a} \right)^{2} + {\Omega}_{\Lambda} 
\end{eqnarray} where ${\Omega}_{i} := {{\rho}_{i}}/{{\rho}_{crit}}$ are the $\Lambda$CDM parameters for radiation (${\rho}_{r}$), matter (${\rho}_{m}$), dark energy (${\rho}_{\Lambda}$) and curvature (${\Omega}_{K} := -K/H_0^2{a}_{o}^2$). Since redshift $z$ depends on the scale factor as $1+z = {a}_{o}/a$, trivially, the parameter Hubble $H$ depends on the redshift $z$ \citep{A6}. International projects assuming the $\Lambda$CDM model, such as the Wilkinson Microwave Anisotropy Probe (WMAP) and the Planck Mission, have estimated that ${\Omega}_{r} \approx 8.4 \cdot 10^{-5}$, ${\Omega}_{m} = 0.3089 \pm 0.0062$, ${\Omega}_{\Lambda} = 0.6911 \pm 0.0062$ and ${\Omega}_{K} \approx 0$. Due to the interest in this work, the parameter of curvature ${\Omega}_{K}$ has not been neglected. 

\subsection{D. Measurement of cosmological distances}
\label{sec:distances}

Some definitions are also required, especially to distinguish among the main cosmological distances: the light-travel distance (or universe age), the comoving distance, the angular distance, the luminosity distance and the distance modulus. 

\begin{figure}
	\includegraphics[width=\columnwidth]{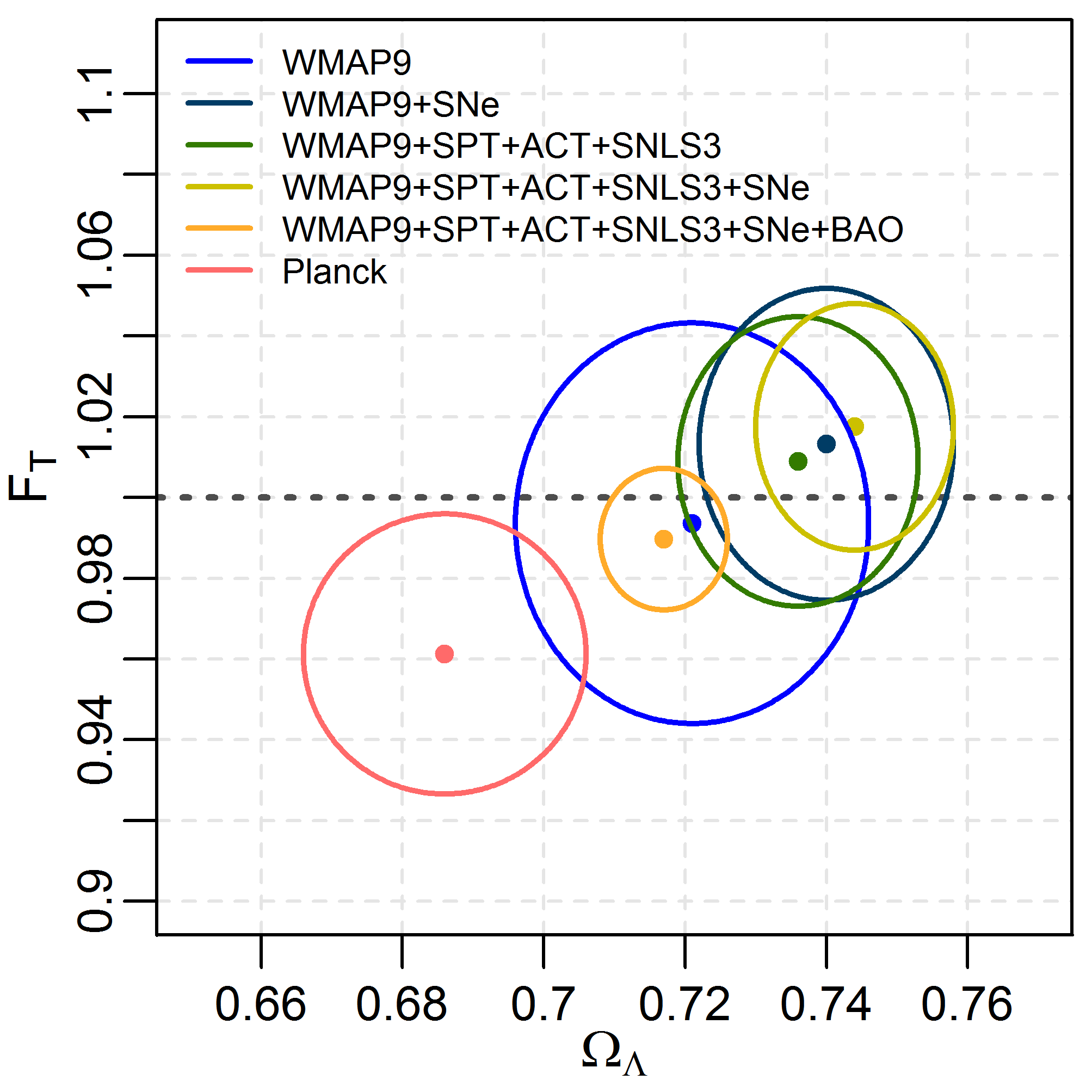} 
\caption{Quotient $F_T$ between the age of universe age ($t_{\hat{o}}$) and the Hubble time ($1/H_o$), according to several studies under the flat standard model (one sigma interval error): 
WMAP9 ($\Omega_{\Lambda} = 0.721 \pm 0.025$), 
WMAP9+SNe ($\Omega_{\Lambda} = 0.740 \pm 0.018$), 
WMAP9+SPT+ACT+SNLS3 ($\Omega_{\Lambda} = 0.736 \pm 0.017$), 
WMAP9+SPT+ACT+SNLS3+SNe ($\Omega_{\Lambda} = 0.744 \pm 0.014$), 
WMAP9+SPT+ACT+SNLS3+SNe+BAO ($\Omega_{\Lambda} = 0.717 \pm 0.009$), 
Planck 2015 ($\Omega_{\Lambda} = 0.686 \pm 0.020$) \citep{A15, B1, B7}. 
}
    \label{fig:1}
\end{figure}

\textit{Light-travel distance}. If the Hubble time is defined as ${t}_{H} = 1/H$, the light-travel distance or lookback time $dt$ is linked with the scale factor $1+z$ as ${dt}/{{t}_{H}} = -{dz}/{(1+z)}$. Therefore, the age of the universe can be obtained from the total light-travel distance \citep{A6}:
\begin{eqnarray} \label{eq:universe.age}   
{t}_{0} = \int^{\infty}_{0} \frac{dz}{(1+z)H(z)} = \frac{1}{H_{0}} {F}_{T} 
\end{eqnarray} where ${F}_{T}$ is the time correction factor, i.e., the quotient between the Hubble time ($H_{0}^{-1}$) and the age of the universe (${t}_{0}$). For the $\Lambda$CDM model, it is:
\begin{eqnarray} \label{eq:factor}   
  {F}_{T} = \int^{\infty}_{0} \frac{{dz}}{(1+z) \sqrt{ {\Omega}_{m} (1+z)^{3} + {\Omega}_{K} (1+z)^{2} + {\Omega}_{\Lambda}}}  
\end{eqnarray} 
Particularly, for a flat universe without any cosmological constant it is obtained $F_T = 2/3$. However, for the WMAP values (${\Omega}_{m}$, ${\Omega}_{\Lambda}$) = (0.279, 0.721), the correction factor is close to 1, and the Planck values (${\Omega}_{m}$, ${\Omega}_{\Lambda}$) = (0.314, 0.686) showed that factor is about $F_T = 0.95$ (Fig.~\ref{fig:1}). This leads to the hypothesis that linear expansion could be compatible with observations within an appropriate theoretical frame (see Sect. II and IV).

\textit{Comoving distance}. Another interesting measurement is the comoving distance $r'$. This is given by the null geodesic with no angular variations ($d \it{\Omega}=0$) in the considered metric. This distance can be written using the redshift $z$ and the Hubble parameter, $H = \dot{a}/a$, with the scale factor expressed as $a/a_{\hat{o}} = 1/(1+z)$ where $a_{\hat{o}} \equiv 1$ is the current value of $a$. For example, using spherical coordinates $(t, r', \theta, \phi)$ in a diagonal metric $g$ = (1, $g_{r'r'}$, $g_{\theta \theta}$, $g_{\phi \phi}$), the comoving distance is given by:
\begin{eqnarray} \label{eq:comoving.distance0}   
  \int^{r'}_{0} \frac{\sqrt{g_{r'r'}} d r'}{a}  = -\int_{t_{\hat{o}}}^{t} \frac{dt}{a} = \int_{0}^{z} \frac{dz}{H(z)}
\end{eqnarray} 
For a $\Lambda$CDM universe, one can find:
\begin{eqnarray} \label{eq:comoving.distance}   
  r' = \sin_K \int^{z}_{0} \frac{d{z}}{H_o \sqrt{ {\Omega}_{m} (1+{z})^{3} + {\Omega}_{K} (1+{z})^{2} + {\Omega}_{\Lambda}}} 
\end{eqnarray} 
where $\sin_K x := \lim_{\epsilon \rightarrow K} \epsilon^{-1/2} \sin (\epsilon^{1/2} x)$, i.e., $\sin_0(x) = x$, $\sin_{+1} x = \sin x$, $\sin_{-1} x = \sinh x$, and recall that  $K = -\Omega_K H_o^2$.

\textit{Angular distance}. Angular distance $r_A$ is given by the proper area of an infinitesimal surface $dA$ with $t$ and $r'$ constant, i.e.: 
\begin{eqnarray}\label{eq:proper.area}  
dA := \sqrt{g_{\theta \theta} g_{\phi \phi}} d\phi  d\theta  :=  r_A^2 \sin \theta d\phi d\theta
\end{eqnarray} For metrics with $g_{\phi \phi} = - a^2{r'}^2 \sin^2 \theta$ and $g_{\theta \theta} = - a^2{r'}^2$, the angular distance is related to the comoving distance as $r_A = {a}{r^{\prime}}$.

\textit{Luminosity distance}. Using the above result, the luminosity distance is $r_L := r_{A}/a^2 = r^{\prime}/a$, according to the Etherington's reciprocity relation \citep{Ellis2007}. Note that inhomogeneous metrics of the Lema{\^\i}tre-Tolman-Bondi (LTB) type also satisfy the same relation between the angular distance and luminosity distance \citep{Yana2015}.

\textit{Distance modulus}. The luminosity distance is empirically measured using the distance modulus ($\mu_{obs}$), defined as the difference between the absolute ($M$) and the apparent ($m$) magnitudes, i.e. $\mu_{obs} := m - M$, but modified by K-correction \citep{A8, A9}:
\begin{eqnarray} \label{eq:luminosity.distance}   
{r}_{L} = 10^{1+\frac{m - M + K_c}{5}}
\end{eqnarray} where $K_c$ is the K-correction, which is dependent on the spectral index $\alpha_{o}$: 
\begin{eqnarray} \label{eq:K.correction}   
K_c(z)= -2.5(1+\alpha_{o})\log_{10}(1+z)
\end{eqnarray} 

The luminosity distance can be estimated using the apparent magnitude of the B-band, remembering that its maximum absolute magnitude is about -19 \citep{A13, A14}. However, it should be noted that K-correction is not unique \citep{A12}. 

\textit{Minimisation problem}. In order to reduce the above uncertainties, distance modulus $\mu_{obs}$ is directly used in minimisation problems of cosmological models:

For a theoretical luminosity distance $r_L$, the modulus is $\mu_{theo} := 5 \log (r_L/ \mathrm{Mpc}) + 25$. The theoretical distance modulus is usually rewritten using the current value of the Hubble parameter $H_o$ as:   
\begin{eqnarray} \label{eq:distance.modulus}  
\mu_{theo} := 5 \log_{10} (r_L H_o) + \tilde{M}
\end{eqnarray} where $\tilde{M}$ can be supposed as a redefined constant, which is degenerate with $H_0$. The value of this constant can be obtained by minimisation, in each case, according to:
\begin{eqnarray} \label{eq:M.modulus}  
 \tilde{M} = \frac{ \displaystyle\sum_{i}{  \frac{\mu_{obs,i} - 5 \log_{10}(r_{L,i} H_o) }{\sigma_{\mu_{obs,i}}^2} } }{\displaystyle\sum_{i}{\sigma_{\mu_{obs,i}}^{-2}}}
\end{eqnarray} where $\sigma_{\mu_{obs,i}}$ is the error interval of each observed modulus $\mu_{obs,i}$. Finally, the Pearson's chi-squared $\chi^2$ is taken as a contrast statistic or measure of discrepancy between the model estimation error and the observed error (sample variance): 
\begin{eqnarray} \label{eq:chi2}  
\chi^2 =  \displaystyle\sum_{i}{  \frac{\left( \mu_{obs,i} - \mu_{theo,i} \right)^2 }{\sigma_{\mu_{obs,i}}^2} }
\end{eqnarray} In order to check the observational compatibility of a proposed model, the predicted distance modulusis compared with the observed one. For example, we can consider Type Ia supernovae from the Supernova Cosmology Project (SCP) Union2.1 database \citep{B10} \citep{Suzuki2012}.

\section{II. General considerations}
\label{sec:Considerations}
\subsection{A. Hypothesis of hyperconical universe}
\label{sec:Hypothesis}

Let $M \subset \mathbb{R}^{1,4}$ be a support manifold contained in the 5-dimensional Minkowski space, more specifically $M := \mathbb{R}_{\geq 0}\times \mathbb{R}^{4}$, and with scalar product $(\cdot)$ given by the most elementary Lorentzian metric $\eta_{1,4}$ of signature (1,4). To restrict the 5-dimensional support manifold $M$ to a 4-dimensional manifold H${}^4$, this paper explores a constraint condition based on a hyperconical universe with linear expansion. Particularly, the proper distance between two points $X,O \in M$ is expanding as a function of an observable parameter time $t \in \mathbb{R}_{\geq 0}$: 
\begin{eqnarray} \label{eq:H}  
 \mathrm{H}^{4} := \left\{ X \in M : \left|X - O \right|_{\eta_{1,4}} = \beta_o t  \right\}
\end{eqnarray} where $\beta_o \in \mathbb{C}$ is a constant respect to the time $t \in \mathbb{R}_{\geq 0}$ but is dependent on the chosen $O \in M$. For simplicity, $O$ is taken as the origin of the universe and $\beta_o$ is assumed as a universal constant. Let $C := (T_{O}M, I_d, \eta_{1,4})$ be a coordinate system or chart such that the coordinates of the points $X,O \in T_{O}M$ are respectively $X = (x^0, ..., x^4)$ and $O = (0,...,0)$, with the identity function $I_d$. For convenience, the coordinates of $X$ are rewritten as $(t_X, \vec{r}, u)$, where $\vec{r} := (x^1, x^2, x^3) \in \mathbb{R}^{3}$ is the ordinary 3-vector, $u := x^4 \in \mathbb{R}$ is the additional spatial dimension and $t_X := x^0 \in \mathbb{R}_{\geq 0}$ is the time dimension. Choosing coordinates such as $t_X = t$, the condition of the hypersurface H${}^4$ is now $\nu^2 t^2 - \vec{r}^2 - u^2 = 0$ with $\nu^2 := 1-\beta_o^2$. Note that the hypercone with $\nu >0$ is an asymptotic limit of hyperboloid manifolds. In fact, if the constraint condition is taken as $t^2 -  \vec{r}^2 - u^2 = \alpha$ with constant $\alpha \neq 0$, the spaces are known as \textit{de Sitter universes} \citep{F1, F2}. Moreover, H${}^4$ is not a Dirac-Milne universe because it admits positive spatial curvature.

The manifold (H${}^4$, $\eta_{1,4}$) is spatially homogeneous and isotropic with respect to $C$. The homogeneity of H${}^4$ is verifiable because it can be foliated by spatial hypersurfaces ${\Sigma_t}$ such that, $\forall p, q  \in {\Sigma_t}$ and $\forall t$, there exists a transformation (diffeomorphism) carrying the point $p$ to point $q$ and leaving the metric invariant. In other words, a spheroidal submanifold ${S_t^3}$ can be defined for each time $t$ as the intersection between the hypercone H${}^4$ and the isochronous hyperplane at this time $t$:
\begin{eqnarray} \label{eq:S3}   
{S_t^3} := \left\{ (\vec{r},u) \in \mathbb{R}^{4} : { \vec{r}}^2 + u^2 = \nu^2 t^2 \right\} \subset \mathrm{H}^4
\end{eqnarray} 
Note that for $\beta_o > 1$, the $u$ component Wick-rotates and $S_t^3$ becomes a 3-hyperboloid (negative curvature). In any case, the time coordinate $t$ is considered as the age of the universe and the $t$-isochronous 3-spheroid or 3-parabloid ${S_t^3}$ could be homeomorph to our expanding spatial universe (Eddington idea), locally conformally flat. Consequently, the universe manifold is globally hyperbolic, i.e, each ${S_t^3}$ is also a Cauchy surface.

In addition, H${}^4$ is spatially isotropic. That is, the manifold H${}^4$ can be covered by a set of timelike curves $\left\{ X|_{\gamma} \right\} \subset$ H${}^4$ and $\forall p \in X|_{\gamma}$ and $\forall v,w \in {T_p}{S_t^3}$ orthogonal to $X|_{\gamma}$ there exists a transformation (diffeomorphism) that, leaving $p$ fixed, carries $v$ to $w$ leaving the metric invariant. These curves $\left\{ X|_{\gamma} \right\}$ are called comoving observers.

The expansion of the universe is an absolute movement with respect to $O = (0,\vec{0},0)_C$, which is fixed (respect to chart $C$). Therefore, the minimum movement of ``particles'' corresponds to the comoving observers, for example $(t,\vec{0},\nu t)_C \in \mathrm{H}^4$. These particles are interpreted as points at rest with respect to the expanding universe ${S_t^3}$.

\subsection{B. Reference of an observer}  
\label{sec:Moving.charts}

According to an \textit{expected equivalence}, an observer that lives in H${}^4$ will measure local distances as in the Minkowskian space $\mathbb{R}^{1,3}_{\eta} := (\mathbb{R}^{4}, \eta_{1,3})$. Therefore, the proper time must be the same. For instance, let $x_0, x_1 \in \mathbb{R}^{1,3}_{\eta}$ be two static points of an observer with coordinates $x_0 = (t_0,\vec{0})$ and $x = (t,\vec{0})$ with $t > t_0 > 0$. Their extended points in H${}^4$ are ${x_0}' = (t_0, \vec{0}, \nu t_0)$ and ${x}' = (t, \vec{0}, \nu t)$, as we can take spaces $\mathbb{R}^{1,3}_{\eta}$ that intersect to H${}^4$ at points $x_0$ and $x$. 

Then, there exists some diffeomorphism $f:(\mathrm{H}^4  \smallsetminus \left\lbrace O \right\rbrace, \eta_{1,4})  \rightarrow (\mathbb{R}_{>0} \times \mathbb{R}^3, g) := \mathbb{R}^{1,3}_{g}$  with $x_0 = f(x_0')$ and $x = f(x')$, such that the metric $g$ inherits properties of $\mathrm{H}^4$ and produces the same proper time $t-t_0$ in $\mathbb{R}^{1,3}_{g}$ as in $\mathbb{R}^{1,3}_{\eta}$:
\begin{eqnarray} \label{eq:distances}   
t - t_0 = |x - x_0|_{\eta_{1,3}} = |f({x}') - f({x_0}')|_{g}
\end{eqnarray}That is, some transformation is required to go from the extrinsic view of the manifold $\mathrm{H}^4 \subset \mathbb{R}^{1,4}_{\eta}$ to the intrinsic view (observer) in $\mathbb{R}^{1,3}_{g}$ under a particular metric $g$.

For instance, we can find a linear transformation $\mathcal{T}_{\lambda}$ acting as $\mathcal{T}_{\lambda}:  (\mathrm{H}^4, \eta_{1,4}) \rightarrow (M, \eta_{1,4}) \subset \mathbb{R}^{1,4}_{\eta}$ such that
\begin{eqnarray} \label{eq:distances1}   
t - t_0 = |f({x}') - f({x_0}')|_{g} = |\mathcal{T}_{\lambda}({x}') - \mathcal{T}_{\lambda}({x_0}')|_{\eta_{1,4}} 
\end{eqnarray} is satisfied.
This implies that ${x_0}'':= \mathcal{T}_{\lambda}({x_0}')$ and ${x}'':= \mathcal{T}_{\lambda}({x}')$ have the same $u$-component to be cancelled, e.g. ${x_0}'' = (t_0, \vec{0}, \nu \lambda)$ and ${x}'' = (t, \vec{0}, \nu \lambda) \in  (M, \eta_{1,4})$, for some $\lambda \in \mathbb{R}$. %

Let $t \in \mathbb{R}_{>0}$ and $s \in \mathbb{R}^4$ be respectively the temporal and spatial components of a path $X(t) = (t, s(t)) \subset \mathrm{H}^4$. $\mathrm{H}^4$

Let $x'(t) = (t, \vec{0}, \nu t) \subset (\mathrm{H}^4, \eta_{1,4})$ be the comoving path (observer) such that ${x_0}' = x'(t_0)$ and $x' = x'(t)$ are the used points in Eq.~\ref{eq:distances1}. Because an observer performs measurements in $t_0$ and $t$, the deformed path $x''(t) = \mathcal{T}_{\lambda}( x'(t) )$ must exist in $\mathrm{H}^4$ during these measurements. Then the deformation must be such as $\lambda = t$, i.e. $\mathcal{T}_{t}$.

Thus, $x'' = x'$ and ${x_0}'' = (t_0,\vec{0}, \nu t) \in \mathbb{R}^{1,4}$. Now, the $u$-component of the difference ${x}'- {x_0}''$ is zero, and the proper time is the same that Eq.~\ref{eq:distances}. Therefore metric $g$ of H${}^4$ is induced by the differential line of ${x}'- {x_0}''$ in $\mathbb{R}^{1,4}_{\eta}$:
\begin{eqnarray} \label{eq:distances2}  
\left| d\left({x} - {x_0})\right) \right|_{g}^2= \left| d\left(f({x}') - f({x_0}')\right) \right|_{g}^2 = \nonumber \\ 
= \left| d\left({x}' - {x_0}''\right)   \right|_{\eta_{1,4}}^2
\end{eqnarray} 

The above discussion (Eq.~\ref{eq:distances}-\ref{eq:distances2}) is equivalent to saying that the observer fixes time coordinate for measurements using a initial time $t_{\hat{o}}$, but it is moving with respect to $O := (0, \vec{0}, 0)$. Thus, its reference line is $\widehat{O} := {x_{\hat{o}}}'' = \mathcal{T}_{t}( x'(t_{\hat{o}}) ) = (t_{\hat{o}},\vec{0}, \nu t) \subset M$ (note that $\widehat{O}$ is generally outside H${}^4$, except at $t = t_{\hat{o}}$). This reference line allows us to define the chart $D := (T_{\hat{o}}{M}, \mathcal{T}_{t}, \eta_{1,4})$.

According to the Eq.~\ref{eq:distances2} the transformation produces a deformation of the distance measurements equivalent to the homomorphism $f: \mathrm{H}^4 \rightarrow \mathbb{R}^{1,3}_g$. 
Thanks to this, the symmetric $\mathrm{H}^4$ becomes an inhomogeneous manifold with metric $g$, i.e. $\mathbb{R}^{1,3}_g$.

If $X_C = (t, \vec{r}, u) \subset \mathrm{H}^4$ is any curve, applying transformation $X_C \rightarrow X_D$ and the constraint condition $\nu^2 t^2 - {\vec{r}}^2 - u^2 = 0$, its coordinates in $M$ under $D$ are:
\begin{eqnarray} \label{eq:coord}   
X_D = \left(t - t_{\hat{o}}, { } \vec{r}, { } \nu t \sqrt{1-\frac{ r^2}{\nu^2 t^2}} - \nu t \right)  
\end{eqnarray} where $r := |\vec{r}|$. Spatial components can be written in spherical coordinates as $\vec{r} = \vec{k_r}\nu t \sin{\gamma}$, where $\vec{k_r} := \vec{r}/r$ is the direction of $\vec{r}$ and $\gamma \in [0, \pi)$ is the angle between the vectors $X-O$ and $\widehat{O}-O$. 
For radial curves $X = X|_{\gamma}$ with a constant angle $\gamma$ and constant direction $\vec{k_r}$ with respect to $\widehat{O}$, it is satisfied that: 
\begin{eqnarray} \label{eq:hubble.law}   
0 = d(\vec{k_r} \sin \gamma ) = d\left( \frac{\vec{r}}{\nu t} \right) =  \frac{d\vec{r}}{\nu t} - \frac{\vec{r}}{\nu^2 t^2}dt 
\end{eqnarray} That is, a Hubble's law ($d\vec{r}/dt = \vec{r}/t$) is obtained as the relationship between the velocity $d\vec{r}/dt$ and the position $\vec{r}$ of the comoving particles. Therefore, we can distinguish between the variation of $\vec{r}$ due to the expansion of the universe and the one due to other causes. In this sense, comoving spatial coordinates ${\vec{r}}^{\prime}$ associated to $\vec{r}$ at time $t$ are defined as those that are equivalent for the time $t_{\hat{o}}$ if the only movement has been given by the expansion:
\begin{eqnarray} \label{eq:comoving}   
{\vec{r}}^{\prime} := \frac{t_{\hat{o}}}{t} \vec{r}
\end{eqnarray} With this, coordinates of any curve $X \subset \mathrm{H}^4$ can be written with respect to $D$ as:
\begin{eqnarray} \label{eq:coordD}   
X_D = \left(t - t_{\hat{o}}, \frac{t}{t_{\hat{o}}}{\vec{r}}^{\prime}, -\nu t \left(1 -  \sqrt{1-\frac{{r^{\prime}}^2}{\nu^2 t_{\hat{o}}^2}} \right) \right)  
\end{eqnarray} 

\subsection{C. Choice of projection}
\label{sec:choice.proj}

The deformation $\mathcal{T}_{t}: \mathrm{H}^4 \rightarrow \mathbb{R}^{1,4}_{\eta}$ leads to a differential line that provides the metric $g$, but the output of this transformation is in $\mathbb{R}^{1,4}_{\eta}$ and still has the unobserved spatial $u$-coordinate. Therefore a map $f^i : \mathbb{R}^{1,4}_{\eta}  \smallsetminus \lbrace O \rbrace \rightarrow  \mathbb{R}^{1,3}_g$ is required to remove $u$, satisfying $f = f^i \circ \mathcal{T}_{t}$, i.e. \begin{center}
	\begin{tikzpicture}[node distance=1.5cm, auto]
	\node (A) {$\mathrm{H}^4$};
	\node (B) [right of=A] {$\mathbb{R}^{1,4}_{\eta}$};
	\node (C) [below of=A] {$\mathbb{R}^{1,3}_g$};
	\draw[->](A) to node {$\mathcal{T}_{t}$}(B);
	\draw[->](A) to node [left] {$f$}(C);
	\draw[->](B) to node [right] {$f^i$}(C); 
	\end{tikzpicture}
\end{center}	
In this work, three possible projection maps $\lbrace f^i \rbrace_{i=0}^2$ are tested. The simplest map $f^0 : (t, \vec{r}, u) = (\hat{t}, \vec{\hat{r}}) \in \mathbb{R}^{1,3}_g$ can be chosen such that $u$ is removed leaving the other coordinates invariant, i.e. $\hat{t}(t,\vec{r}) = t$, $\vec{\hat{r}}(t,\vec{r}) = \vec{r}$, $\forall \vec{r}^2 < \nu^2t^2$ satisfying the constraint condition of H${}^4$. This projection is not surjective since its image is a subset of $\mathbb{R}^{1,3}_g$. The map has inverse only if we limit the domain to the points of an hemisphere of $S_t^3$, i.e. for $\gamma := \tan^{-1}(r/u) \in [0, \pi/2)$. 

It is important to test surjective maps defined for one or both hemispheres, i.e. for domains in $\gamma \in [0, \pi/2)$ and $\gamma \in [0, \pi)$. These maps should be azimuthal and locally conformal projections, that is, $f_{\hat{t}}(t,\vec{0}) \equiv t$, $f_{\hat{r}}(t,\vec{0}) \equiv 0$ and $f_{\hat{r}}(t,\vec{\epsilon}) \approx \vec{\epsilon}$ for $\left| \vec{\epsilon} \right| \ll t$. 

Let $y'(t) := (t, \vec{0}, - \nu t) \ni O$ be antipodal in $S_t^3$ of the comoving observer $x'(t) = (t, \vec{0}, \nu t)$, with $t \in \mathbb{R}_{\geq 0}$. A diffeomorphism $f^1 : \mathrm{H}^4 \smallsetminus \lbrace y' \rbrace  \rightarrow \mathbb{R}^{1,3}_g$ is defined for $\gamma \in [0, \pi)$ by the projection $(t, \vec{r}, u) \rightarrow (f^1_{\hat{t}}(t,\vec{r}), f^1_{\hat{r}}(t,\vec{r})) = (\hat{t}, \vec{\hat{r}}) \in \mathbb{R}^{1,3}_g$ with:
\begin{eqnarray} \label{eq:diffeomorphism}  
f^1_{\hat{r}}(t,\vec{r}) = \nu t \frac{\gamma}{1-\frac{\gamma}{\pi}} \vec{k}_{r}
\end{eqnarray} where $\vec{k}_{r} := \vec{r}/r$ and $\gamma = \gamma(r) := \sin^{-1}(r/\nu^2t^2)$. Finally, another surjective projection map $f^2 : \mathrm{H}^4|_{\gamma < \pi/2} \smallsetminus \lbrace O \rbrace \rightarrow \mathbb{R}^{1,3}_g$ is chosen such as:
\begin{eqnarray} \label{eq:diffeomorphism2}  
f^2_{\hat{r}}(t,\vec{r}) = \nu t \frac{\gamma}{\sqrt{1-\frac{\gamma}{\pi/2}}}\vec{k}_{r}
\end{eqnarray} For simplicity, we suppose that $\hat{t} = f^i_{\hat{t}}(t,\vec{r}) \approx t$ for all considered projections (according to the end of Sect. V.B, this is locally valid).

\section{III. Theoretical features}
\label{sec:Geometry}
\subsection{A. Differential line element and metric tensor}
\label{sec:line.element}

Let $X \subset$ H${}^4$ be any curve. The differential line element $dX$ is easily obtained knowing that $d{\vec{r}}^{\prime}$ can be decomposed in spherical coordinates as $d{\vec{r}}^{\prime} = d{r}^{\prime}\vec{k_r} + r^{\prime}d\Sigma \vec{k_{\Sigma}}$, where $d\Sigma \vec{k_{\Sigma}} := d\theta^{\prime} \vec{k_{\theta}} + \sin \theta^{\prime} d\phi \vec{k_{\phi}}$ is orthogonal to the radial direction $\vec{k_r}$. According to the Eq.~\ref{eq:distances2}, and since $dt_{\hat{o}} = 0$, the square of the differential line element of $X|_D$: 
\begin{eqnarray} \label{eq:coordD1}    
dX^2 = dt^2 \left({2\nu^2} \sqrt{1-\frac{{ r^{\prime}}^2}{\nu^2 t_{\hat{o}}^2}} - {2\nu^2} + 1 \right) 
\nonumber \\ 
-  \frac{t^2}{t_{\hat{o}}^2} \left( \frac{d{r^{\prime}}^2}{1-\frac{{r^{\prime}}^2}{\nu^2 t_{\hat{o}}^2}} + {r^{\prime}}^2d{\Sigma}^2 \right)
-  \frac{{t}}{t_{\hat{o}}} \frac{{ 2r^{\prime}}}{t_{\hat{o}}} \frac{dr^{\prime}dt}{\sqrt{1-\frac{{r^{\prime}}^2}{\nu^2 t_{\hat{o}}^2}}} 
\end{eqnarray} where we identify the curvature $k^{-1} := \nu^2 = 1-\beta_o^2$ and the scale factor $a := t/{t_{\hat{o}}}$. It is easy to identify the elements of the metric tensor $g_{\mu \nu}$ for the universe described by the chart D. Defining $b := \sqrt{1 - k{r^{\prime}}^2/t_{\hat{o}}^2}$, non-zero elements are: 
\begin{eqnarray} \label{eq:elements} 
g_{00} = 2k^{-1}(b-1)+1 \\ 
g_{r'r'} = -\frac{a^2}{b^2} \\
g_{0r'} = -\frac{a r'}{t_{\hat{o}}b} \\
g_{\theta \theta} = -a^2 {r'}^2 \\
g_{\phi \phi} = -a^2{r'}^2 \sin^2 \theta
\end{eqnarray} Note that $g_{00}$ has no problems for the case when $k \rightarrow 0$, since there exists the limit case.  

\begin{figure}
	\includegraphics[width=\columnwidth]{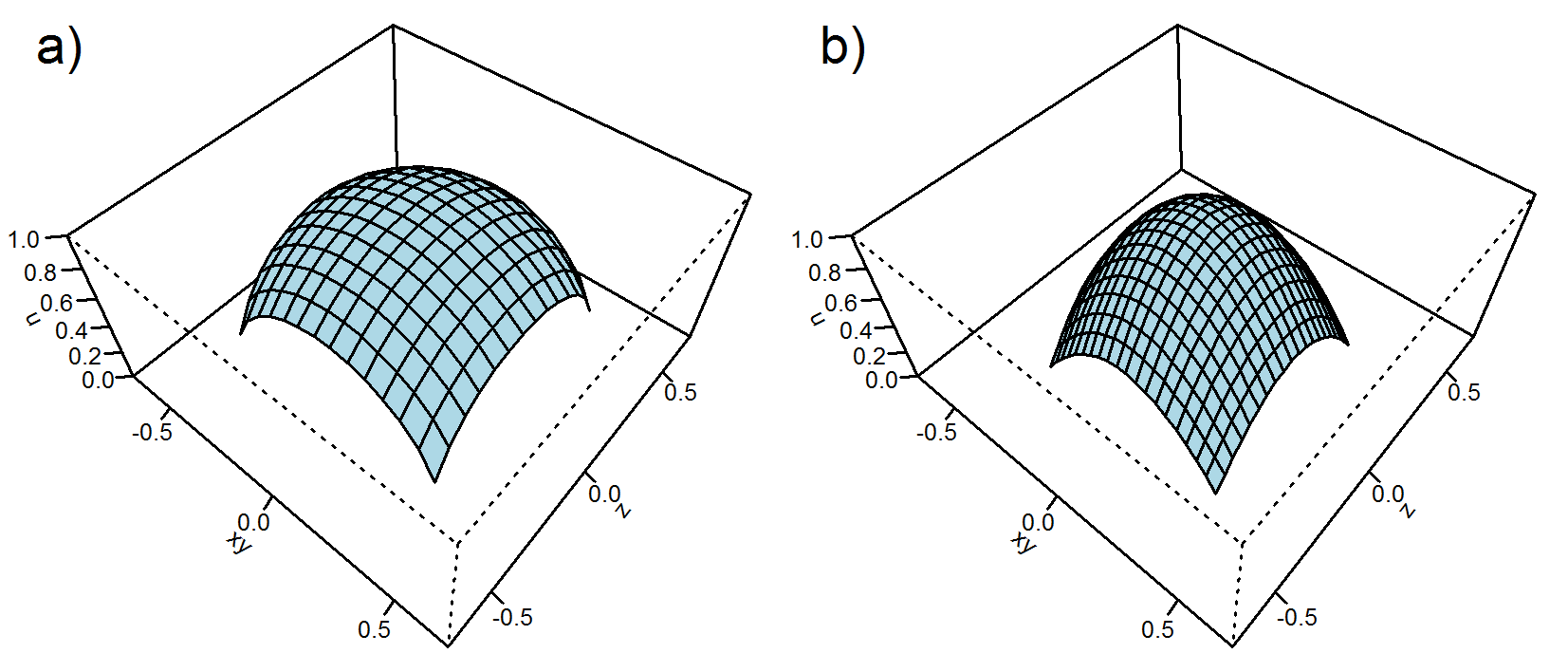} 
\caption{Isochronous hypersurfaces normalised to universe age scale ($t_o \equiv 1$) for:  a) Homogeneous space ($t$-isochronous), obtained considering coordinates such that $g_{0r} \neq 0$ and $g_{00} \neq 1$; b) inhomogeneous space ($t^{\prime}$-isochronous), obtained considering coordinates such that $g_{0r}^{\prime}=0$ and $g_{00}^{\prime}=1$. 
}
    \label{fig:0}
\end{figure}

Regarding chart $D$, some features can be highlighted. Symmetrical spatial elements $g_{ii}$ obtained from Eq.~\ref{eq:coordD1} are compatible with some FRW metrics, but elements $g_{00} \neq 1$ and $g_{0r} \neq 0$, respectively, imply \textit{lapse} and \textit{shift} terms, as in the Arnowitt-Deser-Misner (ADM) formulation of gravity \citep{Deruelle2010}. To compare $g$ with the FRW metrics, a diagonal version of the metric is given by the coordinate change $t^{\prime} := t\sqrt{2k^{-1}(b-1)+1}$, which is equivalent to selecting $g_{00}^{\prime} = 1$, $g_{0r}^{\prime} = 0$ and:
\begin{eqnarray} \label{eq:grr}   
 g_{r'r'}^{\prime} = g_{r'r'} - \frac{g_{0r}^2}{g_{00}} = -{a(t',r')}^2 \frac{1-k^{-1}(b-1)^2}{b^2 2k^{-1}(b-1)+1)} \\ \label{eq:gff}   
 g_{\phi \phi}^{\prime} = -{a(t',r')}^2 {r'}^2 sin^2 \theta \\ \label{eq:ghh}  
 g_{\theta \theta}^{\prime} = -{a(t',r')}^2 {r'}^2              
\end{eqnarray}
where $a(t',r') = t'/(t_{\hat{o}}\sqrt{2k^{-1}(b-1)+1})$. Despite this, differences remain in the FRW metric, resulting in an inhomogeneous hypersurface, i.e. a radially dependent model. Particularly, the $t^{\prime}$-isochronous hypersurface is similar to a paraboloid, in contrast to the homogeneous hypersphere given by the $t$-isochronous (Fig.~\ref{fig:0}). 

Note that this radial inhomogeneity is not a LTB type because the scale factors of angular expansion $a_{\Sigma} := r/r' = a$ and radial expansion $a_{r} := dr/dr' = a$ (see \citep{Yana2015}) do not satisfy the relation $a_{r} = d(a_{\Sigma} r')/dr'$ for the function $a = a(t',r')$ expressed in terms of $t'$ and $r'$.  

\subsection{B. Redshift-distance relation}
\label{sec:Redshift-distance} 

This section seeks to obtain the relation between the comoving distance $r'$, the scale factor $a$ and the redshift $z$. For this purpose, firstly the null geodesic curve is obtained for a photon ($d\it{\Omega} = 0$). Using the diagonal metric $g'$ (Eqs.~\ref{eq:grr}-\ref{eq:ghh}) from the coordinate change $t' := t\sqrt{2k^{-1}(b-1)+1}$ where $b(r') = \sqrt(1 - {k{r'}^2}/t_{\hat{o}}^2)$, the null geodesic curve is:
\begin{eqnarray} \label{eq:geodesic1}   
dt' = - a(t',r') \sqrt{ \frac{1-k^{-1}(1-b)^2}{b^2(2k^{-1}(b-1)+1)} } dr' =
\nonumber \\ 
  - \frac{a}{a_{\hat{o}}} \frac{\sqrt{1-k^{-1}(1-b)^2} }{b(2k^{-1}(b-1)+1)} dr'
\end{eqnarray} where we have defined $a_{\hat{o}}(r') := a(t_{\hat{o}}',r')$ with $t_{\hat{o}}'=t_{\hat{o}}$, i.e., $a_{\hat{o}} = 1/\sqrt{2k^{-1}(b-1)+1}$. Then, the energy $E$ of the photon along the geodesic is described using an affine parameter $s$ and the linear momentum $p := dr'/ds$:
\begin{eqnarray} \label{eq:energy}   
E := \frac{dt'}{ds} = - a \sqrt{ \frac{1-k^{-1}(1-b)^2}{b^2(2k^{-1}(b-1)+1)} } p
\end{eqnarray} In addition, according to the geodesic equation, it is obtained that:
\begin{eqnarray} \label{eq:geodesic2}   
\frac{dE}{ds} = \frac{d^2t'}{{ds}^2} = - \cancelto{0}{\Gamma^{0}_{00}} E^2 - \Gamma^{0}_{r'r'} p^2 - 2\cancelto{0}{\Gamma^{0}_{0r'}} p E =  \nonumber \\
= - \Gamma^{0}_{r'r'} p^2
\end{eqnarray} where $\Gamma^{\alpha}_{\beta\gamma}$ are the Christoffel symbols in coordinates $(t', r', \theta, \phi)$. Particularly, $\Gamma^{0}_{rr} = -\frac{1}{2}{{g'}^{00}} \partial_{t'} g_{r'r'}'$, then:
\begin{eqnarray} \label{eq:geodesic3}   
\frac{dE}{ds} = a\dot{a} \frac{1-k^{-1}(1-b)^2}{b^2(2k^{-1}(b-1)+1)} p^2 = - \frac{\dot{a}}{a} E^2
\end{eqnarray} where $\dot{a} := \partial_{t'}a = a/t' = a_{\hat{o}}/t_{\hat{o}}$. Therefore, dividing Eq.~\ref{eq:geodesic3} and ~\ref{eq:energy} it is obtained that:
\begin{eqnarray} \label{eq:redshift1}   
\frac{dE}{dt'} = \frac{dE/ds}{dt'/ds}= - \frac{\dot{a}}{a} E
\end{eqnarray} Since $H' := \dot{a}/a = 1/t' = (a_{\hat{o}}/a)\partial_{t'}(a/a_{\hat{o}})$ only depends on time $t'$, the integration of Eq.\ref{eq:redshift1} trivially is:
\begin{eqnarray} \label{eq:redshift2}   
\int_{E(t')}^{E_{\hat{o}}} \frac{dE}{E} = - \int_{t'}^{t_{\hat{o}}'} \frac{a_{\hat{o}}}{a} \frac{d}{dt'} \left( \frac{a}{a_{\hat{o}}} \right) dt' \rightarrow \frac{E_{\hat{o}}}{E(t')} = \frac{a}{a_{\hat{o}}}
\end{eqnarray} where $E(t') := 1/\lambda$ and $E_{\hat{o}} := 1/\lambda_{\hat{o}}$ are respectively the emitted and received energy with wavelengths $\lambda$ and $\lambda_{\hat{o}}$. Therefore, redshift is:
\begin{eqnarray} \label{eq:redshift3}   
z :=\frac{\lambda_{\hat{o}}-\lambda}{\lambda} = \frac{a_{\hat{o}}}{a} - 1
\end{eqnarray} and then, $a/a_{\hat{o}} = 1/(1 + z)$. Taking the derivative with respect to $t'$:
\begin{eqnarray} \label{eq:redshift4}   
\frac{d}{dt'}\left( \frac{a}{a_{\hat{o}}} \right) = - \left( \frac{a}{a_{\hat{o}}} \right)^2 \frac{dz}{dt'}
\end{eqnarray} and using the null geodesic curve (Eq.~\ref{eq:geodesic1}):
\begin{eqnarray} \label{eq:redshift5}
\frac{dz}{dr'} = \frac{\sqrt{1-k^{-1}(1-b)^2} }{b(2k^{-1}(b-1)+1)} H'
\end{eqnarray} where $H' = 1/t' = a_{\hat{o}}/(t_{\hat{o}}a) = (1+z)/t_{\hat{o}}$ is a function on $z$. Therefore, the proper distance $r'$ can be obtained from the redshift $z$. Particularly, the comoving distance $r^{\prime}$ is given by:
\begin{eqnarray} \label{eq:comoving.distance2} 
 \xi_{k} \left( \frac{r^{\prime}}{t_{\hat{o}}} \right) := \int_{0}^{r^{\prime}} \frac{\sqrt{1-k^{-1}(1-b)^2}}{b(2k^{-1}(b-1)+1)} \frac{d r'}{t_{\hat{o}}}  =  \int_0^z \frac{d z'}{1 + z}\end{eqnarray} Isolating the comoving distance and applying the map $f^i_{\hat{r}}$ (Sect. II.C),
\begin{eqnarray}\label{eq:comoving.distance3}
r^{\prime} = {t_{\hat{o}}} \xi_{k}^{^{-1}}( \ln(1 + z) ) \\ \label{eq:comoving.distance4}
\hat{r}' = f^i_{\hat{r}}(t_{\hat{o}}, r') 
\end{eqnarray}

Metric tensor $g'$ describes a universe with linear expansion, directly proportional to its age or time, $t$. That is, Hubble parameter is $H'(t') = a^{-1}\partial{a}/\partial{t'} = 1/t'$, which can be written using $a/a_{\hat{o}} = t'/t_{\hat{o}} = 1/(1 + z)$. 

\subsection{C. Ricci curvature}
\label{sec:Ricci.curvature} 

The Ricci curvature tensor can be calculated from the Riemann curvature tensor defined for signature (1,3) as:
\begin{eqnarray} \label{eq:Ricci}   
R_{\alpha \beta} := R_{\alpha \mu \beta}^{\mu} = \partial_{\mu}\Gamma_{\alpha \beta}^{\mu} - \partial_{\beta}\Gamma_{\mu \alpha}^{\mu} + \Gamma_{\rho \mu}^{\mu}\Gamma_{\alpha \beta}^{\rho} - \Gamma_{\rho \beta}^{\mu}\Gamma_{\alpha \mu}^{\rho}
\end{eqnarray} where $\Gamma_{\alpha \beta}^{\mu}$ are the Christoffel symbols, which for the affine connection Levi-Chivita it is satisfied that $\nabla g = 0$ and $g$ is symmetric (without torsion). Therefore:
\begin{eqnarray} \label{eq:Christoffel}   
\Gamma_{\alpha \beta}^{\mu} = \frac{1}{2} g^{\mu \nu}\left(\partial_\alpha g_{\nu \beta} + \partial_\beta g_{\alpha \nu} - \partial_\nu g_{\alpha \beta} \right)
\end{eqnarray}

To obtain shorter expressions, we have worked with $g_{\mu\nu}$ instead of $g_{\mu\nu}'$, which locally coincide. After making the calculations, we can see that the time component of the Ricci tensor is $R_{00} \equiv 0$ for all positions. This contrasts with the FRW metric, in which it is obtained that $R_{00}^{FRW} = -3\ddot{a}/a$. The spatial elements are:
\begin{eqnarray} \label{eq:Spatial.Ricci}
R_{r'r'} = -\left(\frac{\dot{a}}{a}\right)^2 \frac{2k^2}{2b -b^2 + k-1} g_{r'r'}
\\
R_{\theta \theta} =  -\left(\frac{\dot{a}}{a}\right)^2 \frac{(4b-b^2+2k-3)k^2}{(2b-b^2+k-1)^2} g_{\theta \theta}
\\
R_{\phi \phi}     =  -\left(\frac{\dot{a}}{a}\right)^2 \frac{(4b-b^2+2k-3)k^2}{(2b-b^2+k-1)^2} g_{\phi \phi}
\end{eqnarray} 
Therefore, the manifold is only maximally symmetric in the spatial components ($R_{ij} = 1/3\cdot{R}g_{ij}$) at local scale ($r^{\prime}=0$, i.e. $b=1$). In this case, the spatial part of the Ricci tensor $R_{ij}$ approximates to $R_{ij} \approx - 2k(\dot{a}/a)^2g_{ij}$. This contrasts with the FRW result:
\begin{eqnarray} \label{eq:Spatial.Ricci.FRW}
R_{ij}^{FRW} = - \frac{\ddot{a}}{a}g_{ij}^{FRW} - 2\left(\frac{\dot{a}}{a}\right)^2g_{ij}^{FRW} - \frac{2K}{a^2} g_{ij}^{FRW}    
\end{eqnarray}

Finally, the Ricci scalar curvature is obtained from the trace $R := R_{\alpha}^{\alpha}$ as:
\begin{eqnarray} \label{eq:Ricci.scalar}
R = -\frac{6k^2}{t^2} \frac{2b - \frac{1}{3}(2b^2+1) + k-1}{(2b-b^2+k-1)^2}
\end{eqnarray} For the local limit ($r^{\prime} = 0$), the scalar curvature is simplified as $R \approx -6k/t^2 = -6/(\nu t)^2$, as a 3-sphere (of radius $\nu t$) embedded into $\mathrm{R}^{1,4}$. If $k = 1$, the Ricci scalar corresponds to the case of FRW metric with linear scale factor $a = t/t_{\hat{o}}$ and curvature $K = 0$, i.e., this is apparently flat under the FRW view. 

Despite the assumption of linear expansion with constant factor $\beta_o$, this expansion could depend on the spatial coordinates $(\vec{r}, u(\vec{r}))$, of course excluding the time $t$, and then a most general constraint is $\left|X - O \right|_{\eta} = \beta(\vec{r},u) t$. In this case, the resulting curvature $k = 1/(1 - \beta(\vec{r},u)^2)$ also depends on the spatial coordinates. However, we can normalise the local limit as $k(0, u(0)) = 1$.

\section{IV. Observational compatibility}
\label{sec:Discussion} 

\subsection{A. Expansion of the universe}
\label{sec:Expansion} 

The proposed model suggests that the ``radial velocity'' of the expanding $S^3_t$ is constant ($\nu = k^{-1/2}$ does not depend on time $t$) in contrast to the standard cosmology, which describes an acceleration \citep{B9, B11, A14, Suzuki2012}.

The observational compatibility of the hyperconical model was checked using the 580 pairs of redshift and modulus distance from the Union2.1 database \citep{B10} \citep{Suzuki2012}. Results were compared with the standard $\Lambda$CDM cosmology. Their corresponding luminosity distance are:  
\begin{eqnarray} \label{eq:luminosity.dist.LCDM}   
{{r}_{L, \Lambda CDM}} = (1+z) {r'}_{\Lambda CDM} \\
\label{eq:luminosity.dist.conic}   
{{r}_{L, hyp}} = (1+z) f^i_{\hat{r}} \left(t_{\hat{o}}, {{r'}_{hyp}} \right) 
\end{eqnarray} where ${r'}_{\Lambda CDM}$ is given by the Eq.~\ref{eq:comoving.distance}, ${{r'}_{hyp}}$ is given by Eq.~\ref{eq:comoving.distance3} and $f^i_{\hat{r}}$ is the projection (Sect. II.C). The curvature parameter $k$ of the hyperconical model and $\Omega_K$ of the $\Lambda$CDM were fitted from the minimisation of $\chi^2$ for the distance modulus ($\mu_{theo}$) predicted by each model (according to Eq. \ref{eq:distance.modulus}). All error measurements are expressed in terms of one sigma confidence interval, obtained from the variation of $\chi^2$. 

\begin{figure}
	\includegraphics[width=\columnwidth]{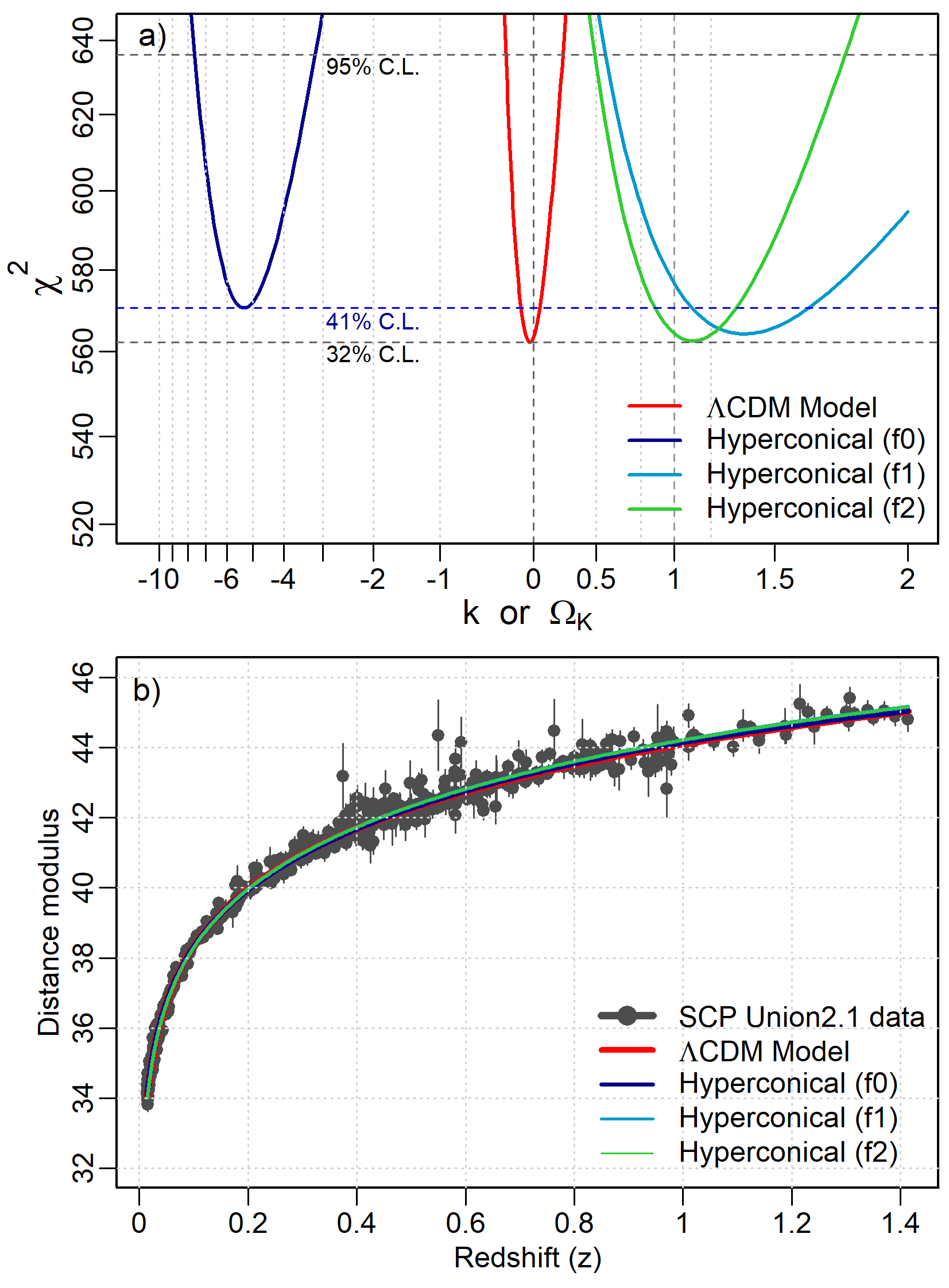} 
\caption{Fitting theoretical distance modulus: $a)$ Model discrepancy ($\chi^2$) depending on curvature, $b)$ Best fit for each model. The used models are the standard $\Lambda$CDM (red line, $\Omega_K = -0.04\pm 0.04$, $\Omega_m \equiv 0.3$, $\Omega_{\Lambda} = 1-\Omega_m-\Omega_K$) and the hyperconical with three projections (dark blue $f^0_{\hat{r}}$, $k = -5.3 \pm 0.5$, light blue $f^1_{\hat{r}}$, $k = 1.36 \pm 0.10$, and green line $f^2_{\hat{r}}$, $k = 1.10 \pm 0.08$) fitted to the distance modulus and redshift from the SCP Union2.1 data.}
    \label{fig:3}
\end{figure}

As a result, the values of $\chi^2$ were minimised up to $\chi_0^2 = 562$ for both the projected hyperconical and $\Lambda$CDM models (Fig.~\ref{fig:3}). The fitted curvature was approximately zero for $\Lambda$CDM ($\Omega_K = -0.04 \pm 0.04$, $\Omega_m \equiv 0.3$, $\Omega_{\Lambda} = 1-\Omega_m-\Omega_K$), but it depends on the considered projection for the hyperconical model. The curvature is negative if the simplest $f^0$ projection is applied ($k = -5.3 \pm 0.5$, $\chi_0^2 = 570$), while it was found positive under the other two projections considered: $f^1$ with $k = 1.36 \pm 0.10$ ( $\chi_0^2 = 564$) and $f^2$ with $k = 1.10 \pm 0.08$ ($\chi_0^2 = 562$). 

Taking into account the 579 degree of freedom, the value of $\chi_0^2 = 562$ corresponds to cumulative probability ($\chi^2 \leq \chi_0^2$) of $33\%$. That is, the value is below the observational limit of $\chi_0^2 = 636$ considered by the confidence level of $95\%$. 

Note that the negative curvature requires a Wick rotation of the $u$-coordinate, low positive values ($0 < k < 1$) imply that $\beta_o$ is imaginary. The obtained curvatures are: $k = 1.36 \pm 0.10$ with $\beta_o^2 = 0.26 \pm 0.06$ and $k = 1.10 \pm 0.08$ with $\beta_o^2 = 0.09 \pm 0.06$. If we consider the normalisation $k := 1$, this corresponds to $\beta_o = 0$. 

On the other hand, the age of the universe is also analysed. As the model proposed describes a universe that expands linearly with time, there is an equivalence between the Hubble time and the age of the universe ($F_T = 1$, according to Eq.~\ref{eq:universe.age}). To contrast this hypothesis, it is required to obtain model-independent measurements of the expansion of the universe and the age. For instance, globular clusters can be used to estimate the lower limit of the age of the universe, but high uncertainty is found for the distances, and thus for the brightness, mass and age. With this, the oldest known globular clusters have stars with $1.4 \pm 0.2 \cdot 10^{10}$ years \citep{KraussChaboyer2003}. This value is compatible with the Hubble time under the hyperconical model ($1.41 \pm 0.04 \cdot 10^{10}$ years), estimated assuming B-band absolute magnitude about $-19.0 \pm 0.5$ for 175 nearby SNe Ia ($z < 0.1$). 

Within the $\Lambda CDM$ frame, the factor $F_T$ is close to one. For the WMAP values ($\Omega_m$, $\Omega_{\Lambda}$) = ($0.266$, $0.732$), this is $F_T = 0.996$, and for the Planck values ($\Omega_m$, $\Omega_{\Lambda}$) = ($0.3089$, $0.6911$), the factor is about $F_T = 0.956$. The analysis of the Planck Collaboration made different assumptions from those used by the WMAP team \citep{B8, A15}. 

Regarding the local-to-global ratio ($H_{0(L)}/H_{0(G)}$) of the Hubble parameter fitted to the SCP SNe Ia data \citep{B5}, the best fit when $H_{0(L)}/H_{0(G)} = 1$ is the well-known $\Omega_M = 0.28$, $\Omega_{\Lambda} = 0.72$ cosmology found by \citet{B6}. These values coincide with the ones which lead to the $F_T$ close to one. That is, the universe has expanded so that just today $H_o = 1/t_o$, although standard theory denies the possibility that $H = 1/t$. With the model proposed in this paper, the coincidence of $F_T \approx 1$ could be no accident.

\subsection{B. Friedmann equations compatibility}
\label{sec:Friedmann.equations} 

Friedmann equations are the result of two assumptions. First, Einstein field equations are applied at the cosmological scale; but perhaps they are only valid locally because they describe the intrinsic curvature caused by a certain distribution of energy density. Second, a perfect fluid stress-energy tensor is considered for the Einstein's equations, leading to homogeneous and isotropic solutions (FRW metric). Therefore, Friedmann equations are obtained assuming that spatial differential $d\vec{\ell}$ of the universe expands with a unknown scale factor $a(t)$ as $d\vec{\ell} = a(t)d\vec{\ell}'$, and then possible solutions of $a(t)$ are sought. In contrast, this study starts from a space-time that expands as $\vec{\ell} = a(t)\vec{\ell}'$ with a known $a(t) = t/t_{\hat{o}}$, i.e. the expansion does not depend on matter content. However, GR must be satisfied at least locally and then possible solutions of the energy distribution can be sought.

Taking the Einstein tensor as $G_{\alpha \beta} := R_{\alpha \beta}  - g_{\alpha \beta} R/2$, and assuming Einstein field equations are valid at least at local scale,
\begin{eqnarray} \label{eq:einstein.field}   
G_{\alpha \beta} - \Lambda g_{\alpha \beta} = 8\pi \mathrm{G} T_{\alpha \beta}
\end{eqnarray} where $\mathrm{G}$ is the gravitational constant, $\Lambda$ is the cosmological constant and $T_{\alpha \beta}$ is the stress-energy tensor. Since $R_{00} = 0$ and locally $R \approx - 6k(\dot{a}/a)^2$, the temporal component of the Einstein tensor is $G_{00} \approx 3k(\dot{a}/a)^2$. And since locally $R_{ij} \approx Rg_{ij}/3$, spatial components are $G_{ij} \approx g_{ij}k(\dot{a}/a)^2$. For convenience, it is considered the local normalisation $k_o := k(0, u(0)) = 1$.

Assuming that the universe is formed by a perfect fluid with pressure $p$ and density $\rho$, its stress-energy tensor is $T_{\alpha \beta} = (\rho + p)u_{\alpha}u_{\beta} - pg_{\alpha \beta}$. If the fluid is at rest, $u_{\alpha} = \delta_{\alpha}^0$, the Einstein equations are locally:
\begin{eqnarray} 
\label{eq:einstein.tt}   
3k_o\left(\frac{\dot{a}}{a} \right)^2 - \Lambda = 8\pi \mathrm{G} \rho
\\
\label{eq:einstein.rr}   
k_o\left(\frac{\dot{a}}{a} \right)^2 g_{ij} - \Lambda g_{ij}  = - 8\pi \mathrm{G} p g_{ij} 
\end{eqnarray} 

We can interpret that universe energies ($\mathrm{G} \rho$ and $\Lambda$) are implicitly defined in Eq.~\ref{eq:einstein.tt} and~\ref{eq:einstein.rr} at the local scale. That is, the energy distribution of the universe can be obtained as a solution of these equations. The equations can be rewritten as: 
\begin{eqnarray} \label{eq:friedmann.1}   
 k_o \left(\frac{\dot{a}}{a} \right)^2 = \frac{8\pi \mathrm{G}}{3} \rho + \frac{\Lambda}{3}
\\ 
\label{eq:friedmann.2}   
 k_o \left(\frac{\dot{a}}{a} \right)^2  = - {8\pi \mathrm{G}} p + {\Lambda}
\end{eqnarray}

Comparing Eq.~\ref{eq:friedmann.1} and \ref{eq:friedmann.2} with the Friedmann equations obtained in the FRW theory (Eq.~\ref{eq:Friedmann.eq1} and \ref{eq:Friedmann.eq2}), we can see that they are compatible for $\ddot{a} = 0$ and $K_{FRW} = 0$, i.e., an apparently flat universe. In fact, Eq. \ref{eq:friedmann.1} can be rewritten as $\rho_{crit}(t) = \rho + \rho_{\Lambda}$, where $\rho_{crit}(t) := {3k_o{H}^{2}}/{8\pi G}$ and ${\rho}_{\Lambda} := {\Lambda}/{8\pi G}$, in agreement with Eq. \ref{eq:Friedmann.eq1b}. That is, critical density is equal to the total universe density (sum of matter density $\rho$ and dark energy density $\rho_{\Lambda}$) at any time $t$. The coincidence between both densities seems to agree with observations, and it is an additional coincidence to the observed time quotient $F_T \approx 1$.

Since $a = t/t_{\hat{o}}$, and defining $w := p/{\rho}$ as the parameter of the matter state equation, it is found that Eq.~\ref{eq:friedmann.1} and Eq.~\ref{eq:friedmann.2} are compatible if and only if $w \neq -1$. A solution for $\Lambda$ and G$\rho$ is given by:
\begin{eqnarray} \label{eq:rho.values}   
  4\pi \mathrm{G}\rho(1+w) = \frac{k_o}{t^2}
\\
\label{eq:lambda.values}   
  \Lambda t^2 = k_o \frac{1+3w}{1+w}
\end{eqnarray}

If $\Lambda$ is finite and constant, there are two possibilities:
\begin{enumerate}
	\item In the limit as $t$ approaches 0, $w$ approaches $-1/3$. 
	\item If $\Lambda \neq 0$, it is expected that $\Lambda t^2 = 3$ for a certain $t$, causing that $w \rightarrow \infty$ and this is impossible.
\end{enumerate}

As $\Lambda$ is constant, it necessarily must satisfy that $\Lambda = 0$ and $w = -1/3$ for any time $t$. This value is compatible with the dark matter equation of state obtained from the rotational curves of galaxies in \citep{C1, C2}. However, the proposed model must be checked with other observables, especially from CMB and kSZ cluster data, which exclude inhomogeneous models as the LTB cosmologies at 1Gpc scale \citep{Bulletal2012}.

\section{V. Interpretation of the model}
\label{sec:Interpretation} 

\subsection{A. Obtaining the metric}\label{obtaining.metric} 

The metric presented in this work cannot be obtained from FRW universes because these combine a static spatial manifold (the most general case of homogeneous and isotropic) and an arbitrary scale factor that depends on time $t$. The scale factor is not derived from the spatial manifold but from the Friedmann equations, i.e. it depends on the matter content according to the Einstein field equations. In contrast, the proposed metric is directly obtained from an embedded manifold, which is homogeneous and isotropic according to a static reference system (chart $C$). The scale factor is directly derived from the expansion of the spatial submanifold and does not depend on the matter contents of the universe (i.e. gravity only perturbs the universe metric, without modifying the global curvature).

The main difference of both procedures is that FRW is obtained using an intrinsic view of the space-time while the proposed metric is obtained from an extrinsic view, i.e. the space-time is embedded into a higher dimensional manifold. The key is to find an adequate transformation from extrinsic to intrinsic view, satisfying the equivalence of proper time measurements. Particularly, it is used a deformation $\mathcal{T}_t$ that expands the spatial section up to the current time $t$ to measure a correct proper time (of a comoving path). Note that, if $\mathcal{T}_{t_0}$ is used instead of $\mathcal{T}_t$, the proper time is also correct but Eq.~\ref{eq:coordD1} becomes a static FRW metric (without scale factor, and then it is an unrealistic universe).

In addition, to obtain a critical density $\rho_{crit}$ compatible with observations (Sect. IV.B), a local normalisation of the curvature is required as $k = 1$. If $\beta$ is chosen as a constant for all spatial positions, it must be $\beta = 0$. For this purpose, we need to find an adequate projection $f^i$ that satisfies $k = 1$ (e.g. $f^2$ is compatible within the two sigma confidence region). In other words, the problem of fitting the proposed model is reduced to find an adequate transformation from the extrinsic view of a symmetric manifold to the intrinsic view of an inhomogeneous manifold.

Note that, for the extrinsic view of the manifold $\mathrm{H}^4$, it is required to consider a centre or focus $O$ from which we can define the hypersurfaces. The centre $O$ is in the support manifold $(M, \eta_{1,4})$, where $(\mathrm{H}^4, \eta_{1,4})$ is embedded. This centre is not in the hypersurface $S_t^3$ for $t > 0$, and therefore it does not exist now. In addition, if we draw the comoving paths corresponding to the points of $S_t^3$, all of them would have the point $O$ at $t = 0$, corresponding to the Big Bang. It is therefore impossible to distinguish points privileged in $S_{t>0}^3$ (Copernican principle).

\subsection{B. Modified Lagrangian density}\label{modified.lagrangian} 

If we assume that expansion of the universe does not depend on matter content, we can think that GR could be invalid at the cosmological scale. Applying the Friedmann equations at local scale, it is obtained that the expansion is only compatible with the equation of state $w = -1/3$ for the energy density of the universe, i.e., $\rho_o = \rho_{crit}$ and $p_o = -\rho_{crit}/3$. This implies that the Einstein field equations can be written as: 
\begin{eqnarray} \label{eq:einstein.field2}   
G_{\alpha \beta}  = 8\pi \mathrm{G} T_{\alpha \beta} = 8\pi \mathrm{G} \left( {T^o}_{\alpha \beta} + {T^m}_{\alpha \beta}\right) 
\end{eqnarray} where ${T^m}_{\alpha \beta}$ is the stress-energy tensor of the ordinary matter-energy and ${T^o}_{\alpha \beta} = (\rho_o + p_o)u_{\alpha}u_{\beta} - p_og_{\alpha \beta}$ is the stress-energy tensor of the background space-time (universe), with $w = -1/3$ and energy density equal to $\rho_{crit} = 3k_o/8\pi\mathrm{G}t^2$. That is, the total Lagrangian density is locally equal to:
\begin{eqnarray} \label{eq:space.lagrangian} 
\mathcal{L} \approx \frac{1}{16\pi\mathrm{G}}\left(R + \frac{6k_o}{t^2} \right) + \mathcal{L_M}   
\end{eqnarray} where $\mathcal{L_M}$ is the Lagrangian term of the ordinary mass-energy. Note that the curvature term $R + {6k_o}/{t^2}$ corresponds to the local difference between the total Ricci curvature $R$ and the local limit of the Ricci curvature estimated for the empty hyperconical universe, $R_u \approx -{6k_o}/{t^2}$. Therefore, a \textit{modified gravity} is required for the general case. The simplest modification for the Lagrangian density is given by the general difference $\Delta R := R - R_u$, that is:
\begin{eqnarray} \label{eq:space.lagrangian2} 
\mathcal{L} = \frac{\Delta R}{16\pi\mathrm{G}} + \mathcal{L_M}   
\end{eqnarray} Therefore, it is a simple type of modified gravity Lagrangian density, which leads to equations similar to those obtained using a flat FRW universe, because the curvature of the universe is subtracted. 

As the linear expansion of the proposed universe does not depend on the matter content, Einstein field equations can be applied like in a flat universe with total compensation of the matter effect (deceleration) by a \textit{dark energy effect} (acceleration). And finally, the local gravity becomes a perturbation theory respect to the (independent) universe metric. 

In order to probe this interpretation, we can derive a perturbation metric $\hat{g}$ from the curvature anomaly $\hat{R} := \Delta R$ of the universe, with all the local symmetries given by a flat FRW metric, i.e.:
\begin{eqnarray} \label{eq:perturbed.metric} 
d{\hat{X}}^2 = \hat{g}_{\mu \nu}{\hat{X}}^{\mu}{\hat{X}}^{\nu} = d\hat{t}^2 - {\hat{a}(\hat{t})} \left( {d{\hat{r}{}'}^2} + {\hat{r}{}'}^2d{\Sigma}^2 \right)   
\end{eqnarray} where the measurements locally coincide for the time coordinate $d\hat{t} = df^i_{\hat{t}}(t, \vec{0}) \equiv dt$ and for the radial coordinates $d\vec{\hat{r}}{}' = df^i_{\hat{r}}(t, \vec{0}) \equiv  d\vec{r}'$. As $\hat{r}{}'$ is theoretically described by the $\Lambda$CDM model, this should be equal to Eq.~\ref{eq:comoving.distance4}. That is, the standard model should be obtained from the local perturbation analysis under the hyperconical model.

Taking the null geodesic curve from the local metric (Eq.~\ref{eq:perturbed.metric}), it is very easy to obtain the expression for the (FRW) comoving distance (Eq.~\ref{eq:comoving.distance0}). And equating with Eq.~\ref{eq:comoving.distance4}, it is obtained the following function composition: 
\begin{eqnarray} \label{eq:igualdad}
\int_0^z dz' \frac{1}{\hat{H}} = f^i_{\hat{r}} \circ \xi_k^{-1} \circ \int_0^z dz' \frac{1}{H}
\end{eqnarray} where $\hat{H} := \hat{a}^{-1} d{\hat{a}}/d\hat{t}$ is the local Hubble parameter and $H = 1+z$ is the global value taking $t_{\hat{o}} \equiv 1$. Therefore, the local expansion is given by:
\begin{eqnarray} \label{eq:local.hubble}
\hat{H}(z) = \left( \frac{d}{dz'} \circ f^i_{\hat{r}} \circ \xi_k^{-1} \circ \int_0^{z'} dz \frac{1}{H}  \right)^{-1} (z)
\end{eqnarray} On the other hand, the first Friedmann Equation is valid at least for the local metric $\hat{g}$ and then, we can identify cosmological parameters equivalent to the radiation content ($\Omega_4$), matter content ($\Omega_3$) and dark energy ($\Omega_1$) for compensation of the local gravity, i.e.
\begin{eqnarray} \label{eq:local.friedmann}
\hat{H}(z) = \hat{H}_0 \sqrt{ \Omega_4 (1+z)^4 + \Omega_3 (1+z)^3 + \Omega_1 }
\end{eqnarray} where $\hat{H}_0 := \hat{H}(0)$. The values of these cosmological parameters can be obtained fitting Eq.~\ref{eq:local.friedmann} to Eq.~\ref{eq:local.hubble}, but they depend on the curvature $k$, i.e. on the projection (Sect. II.C) applied to the function $\xi^{-1}_k$. At least for $z \leq 1.41$, the projection $f^1_{\hat{r}}$ (with $k = 0.45 \pm 0.15$) obtains $\Omega_4 = 0.03 \pm 0.03$, $\Omega_3 = 0.23 \pm 0.03$ and $\Omega_1 = 0.74 \pm 0.01$. For the projection $f^2_{\hat{r}}$ (with $k = 0.80 \pm 0.15$), it is obtained $\Omega_4 = 0.02 \pm 0.02$, $\Omega_3 = 0.25 \pm 0.02$ and $\Omega_1 = 0.73 \pm 0.01$. Note that are similar to the values obtained by WMAP ($\Omega_m = 0.279$, $\Omega_{\Lambda} = 0.721$) under the $\Lambda$CDM model, but they disagree with the Planck results \cite{A3,A15}. 

Moreover, the intrinsic coordinate time $\hat{t}$ can be measured as the light-travel distance (Eq.~\ref{eq:universe.age}), that is $d \hat{t} = -dz/((1+z)\hat{H}(z)) \neq dt$ using Eq.~\ref{eq:local.hubble}. It leads to the identity $\hat{t}(t,\vec{0}) \equiv t$ and the local approximation $\hat{t}(t,\vec{\epsilon}) \approx t$ for $\left| \vec{\epsilon} \right| \ll t$. 

Summarising, $\Lambda$CDM model is an intrinsic view of the universe, with regional coordinates ($\vec{\hat{r}}$,$\hat{t}$) that show an apparent acceleration. The hyperconical model is based on the extrinsic view of a linearly expanding universe that, with an appropriate transformation of the coordinates, is regionally compatible with the $\Lambda$CDM model. An assumption required is that gravity only affects the universe on a local scale, which is equivalent to consider an apparent flat universe. In other words, the matter perturbs the metric but does not affect to the global curvature $R_u \approx -6k(\dot{a}/a)^2$. 

With all this, we see that is possible to built a manifold with positive global curvature and linear expansion (for the local time $t$) that could explain the ``flatness problem'' and the (apparent) existence of the dark energy to compensate the (apparent) action of the cosmic gravity.

\section{VI. Conclusions}
\label{sec:Conclusions} 

The FRW metric used in the standard $\Lambda$CDM model is built combining a universe with homogeneous and isotropic spatial section and a scale factor that is added to explain its expansion. In this paper, the metric is directly obtained for an embedded universe with linear expansion and independent of the matter contents. Particularly, it is considered that any observer is placed in a 3-spheroid or 3-hyperboloid with radial expansion (4-hypercone). One of the key assumptions is that the proper time of each observer is preserved and thus fixes an initial reference time, although its actual position is moving in time due to the expansion.

From this hypothesis, an inhomogeneous metric with increasing curvature radius is obtained. For a particular observer the scalar curvature is $R = -6k/t^2$, as is expected for a 3-spherical surface embedded into $\mathrm{R}^{1,4}$. Taking the normalisation $k := 1$, this Ricci scalar coincides for a flat FRW metric with linear expansion $a = t/t_{\hat{o}}$. It has the same effect as if there is no curvature. This effect implies that total density of the universe is equal to the critical density at any time (``flatness problem''). 

Regarding the observational compatibility, the proposed model leads to an identity between the age of the universe and Hubble time, consistent with a semi-empirical quotient close to one within the standard $\Lambda$CDM frame. Without curvature, the standard model has a unique solution that satisfies this entity, which is given by $\Omega_{\Lambda} = 0.73$ and $\Omega_{m} = 0.27$. For these values, a homogeneous local-to-global ratio ($H_{0(L)}/H_{0(G)} = 1$) is obtained. Expansion of the universe is also compatible for the hyperconical model, which obtains a $\chi_0^2 = 562$, the same value that the standard $\Lambda$CDM model.

If we assume that Einstein field equations are valid at least locally, the hyperconical universe model shows a $t^{-2}$ dependence in all $\Omega$ terms of the Friedmann equations. With this, matter would have an equation of state such that $w = -1/3$ with G and $\Lambda$ constants. To be consistent, this leads to a modified Lagrangian density of the universe, subtracting the term of cosmological curvature. In other words, the gravity becomes a perturbation theory respect a universe expanding regardless of the matter content. With this, the $\Lambda$CDM model can be interpreted as a local perturbation theory compatible with the hyperconical model. 

The two coincidences of the apparent flatness and the age of the universe should not be understood as a mere fortuity that has just happened ``today''. The proposed model in this work is still a first approximation, but it should be understood as a suggestion to be analysed in greater depth. Particularly, other coordinate transformations should be explored. In addition, it must be checked with other observables, especially from CMB and kSZ cluster data, which exclude LTB models at 1Gpc scale.

\section*{Acknowledgements}

\input acknowledgement.tex   


\end{document}

%% file: author_list.tex
\author{R.~Monjo} 
\affiliation{Department of Geometry and Topology, Faculty of Mathematics, Complutense University of Madrid, Spain.\\e-mail: rmoncho@ucm.es}

%% file: acknowledgement.tex
The main idea of the paper emerged from a conversation with my colleague Conrad Gallent. Thank you for encouraging me to develop it. I would also like to thank Dar\'\i{}o Redolat for leading me to continue during this year. All comments of Prof. Antonio L. Maroto and the PRD reviewers have been very helpful to improve this paper.

%% file: Monjo_2017_inhomogeneous_cosmology_v4.bbl
\begin{thebibliography}{99}

  \bibitem[\protect\citeauthoryear{Eddington}{1933}]{RefJ}
A.S.~Eddington,
The Expanding Universe. Cambridge University Press, London. (1933).

  \bibitem[\protect\citeauthoryear{Penrose}{2004}]{RefB}
R.~Penrose,
The Road to Reality: A Complete Guide to the Laws of the Universe. Vintage Books, London. (2004).

\bibitem[\protect\citeauthoryear{Steinhardt}{2011}]{RefC}
P.J.~Steinhardt,
The Inflation Debate. Scientific American {\bf 304}, 36-43 (2011).

\bibitem[\protect\citeauthoryear{Earman \& Moster\'\i{}n}{1999}]{A1}
J.~Earman and J.~Moster\'\i{}n,
Philosophy of Science {\bf 66}, 1-49 (1999).	

\bibitem[\protect\citeauthoryear{Hollands \& Wald}{2002}]{A2}
S.~Hollands and R.M.~Wald,
Gen. Relativ. Gravit. {\bf 34}, 2043-2055 (2002).

\bibitem[\protect\citeauthoryear{Spergel et al.}{2007}]{A3}
D.N.~Spergel {\sl et al.} (WMAP collaboration),
Astrophys. J. Suppl.{\bf 170}, 377-408 (2007). 

\bibitem[\protect\citeauthoryear{Clifton et al.}{2012}]{Clifton1}
T.~Clifton {\sl et al.},
Phys. Rep. {\bf 513}, 1-189 (2012). 

\bibitem[\protect\citeauthoryear{Koyama}{2016}]{Koyama1}
K.~Koyama,
Rep. Prog. Phys. {\bf 79}, 046902 (2016). 

\bibitem[\protect\citeauthoryear{Benoit-L\'evy \& Chardin}{2012}]{Milne1}
A.~Benoit-L\'evy and G.~Chardin,
Astron. Astrophys. 537, id.A78 (2012). doi: 10.1051/0004-6361/201016103

\bibitem[\protect\citeauthoryear{Melia}{2007}]{Melia1}
F.~Melia,
Mon. Not. R. Astron. Soc. 382, 1917-21 (2007). 

\bibitem[\protect\citeauthoryear{Perivolaropoulos et al.}{2009}]{A4}
L.~Perivolaropoulos and A.~Shafieloo,
Phys. Rev. D: Particles and fields 79 (2009). 

\bibitem[\protect\citeauthoryear{Verde et al.}{2013}]{A5}
L.~Verde, R.~Jimenez and S.~Feeney,
Phys. Dark Universe 2, 65-71 (2013). 

  \bibitem[\protect\citeauthoryear{Bennett et al.}{2014}]{B4}
C.L.~Bennett, D.~Larson, J.L.~Weiland and G.~Hinshaw,
Astrophys. J. {\bf 794}, 135 (2014).	

\bibitem[\protect\citeauthoryear{Lima}{2007}]{Lima1}
J.A.S.~Lima,
arXiv:0708.3414 (2007).

\bibitem[\protect\citeauthoryear{Melia \& Abdelqader}{2009}]{Melia2}
F.~Melia and M.~Abdelqader,
Int. J. Mod. Phys. D {\bf 18}, 1889 (2009).

\bibitem[\protect\citeauthoryear{Melia \& Shevchuk}{2012}]{Melia3}
F.~Melia and A.S.H.~Shevchuk,
The Rh = ct Universe. Mon. Not. R. Astron. Soc. {\bf 419}, 2579 (2012).

\bibitem[\protect\citeauthoryear{Lewis}{2013}]{Lewis1}
G.F.~Lewis,
Mon. Not. R. Astron. Soc. {\bf 432}, 2324-30 (2013). 

\bibitem[\protect\citeauthoryear{Jimenez et al.}{2009}]{Jimenez2009} 
J.B.~Jimenez, R.~Lazkoz and A.L.~Maroto,
Phys. Rev. D {\bf 80}, 023004 (2009).

\bibitem[\protect\citeauthoryear{Liddle}{2003}]{A6}
A.R.~Liddle,
An Introduction to Modern Cosmology. Wiley, Chichester (2003).

\bibitem[\protect\citeauthoryear{Planck Collaboration}{2015}]{A15}
Planck Collaboration,
Astron. Astrophys., (2015). doi: 10.1051/0004-6361/201525820. arXiv:1502.01589.

\bibitem[\protect\citeauthoryear{Spergel et al.}{2007}]{B1}
D.N.~Spergel {\sl et al.} (WMAP collaboration),
Astrophys. J. Suppl.{\bf 170}, 377 (2007).

\bibitem[\protect\citeauthoryear{Hinshaw et al.}{2013}]{B7}
G.~Hinshaw {\sl et al.},
Astrophys. J. Suppl.{\bf 208}, 19-44 (2013).

\bibitem[\protect\citeauthoryear{Ellis}{2007}]{Ellis2007}
G.F.R.~Ellis,
Gen. Relativ. Gravit. {\bf 39}, 1047-1052 (2007).

\bibitem[\protect\citeauthoryear{Yana et al.}{2015}]{Yana2015}
X.-P.~Yan, D.-Z.~Liu and H.~Wei,
Phys. Lett. B {\bf 742}, 149–159 (2015). 

\bibitem[\protect\citeauthoryear{Sanders et al.}{1989}]{A8}
D.B.~Sanders, E.S.~Phinney, G.~Neugebauer, B.T.~Soifer and K.~Matthews,
Astrophys. J. {\bf 347}, 29 (1989).

\bibitem[\protect\citeauthoryear{Kennefick \& Bursick}{2008}]{A9}
J.~Kennefick and S.~Bursick,
Astron. J. {\bf 136}, 1799-1809 (2008).

\bibitem[\protect\citeauthoryear{Natali et al.}{1997}]{A12}
F.~Natali, E.~Giallongo, S.~Cristiani and F.~La-Franca,
Astron. J. {\bf 115}, 397-404 (1997).

\bibitem[\protect\citeauthoryear{Burns et al.}{2010}]{A13}
C.R.~Burns, M.~Stritzinger and M.M.~Phillips,
Astron. J. {\bf 1010}, 4040 (2010).	

\bibitem[\protect\citeauthoryear{Conley et al.}{2011}]{A14}
A.~Conley {\sl et al.},
Astrophys. J. Suppl.{\bf 192} (2011). 

\bibitem[\protect\citeauthoryear{Kowalski et al.}{2008}]{B10}
D.R.~Kowalski {\sl et al.},
Astron. J. {\bf 686}, 749-778 (2008).

\bibitem[\protect\citeauthoryear{Suzuki et al.}{2012}]{Suzuki2012}
N.~Suzuki {\sl et al.},
Astrophys. J. {\bf 746}, 85 (2012). 

\bibitem[\protect\citeauthoryear{Wise}{2015}]{F1}
D.K.~Wise,
Theoretical and Mathematical Physics {\bf 19}, 1017-1041 (2015). 

\bibitem[\protect\citeauthoryear{Gryb}{2015}]{F2}
S.~Gryb,
Gen. Relativ. Gravit. {\bf 47}, 37 (2015). 

\bibitem[\protect\citeauthoryear{Deruelle et al.}{2010}]{Deruelle2010}
N.~Deruelle, M.~Sasaki, Y.~Sendouda, D.~Yamauchi,
Progress of Theoretical Physics {\bf 123}, 169-185 (2010). 

\bibitem[\protect\citeauthoryear{Krauss and Chaboyer}{2003}]{KraussChaboyer2003}
L.M.~Krauss and B.~Chaboyer 
Science {\bf 299}, 65-69 (2003). 

\bibitem[\protect\citeauthoryear{Pisantia et al.}{2013}]{B8}
O.~Pisantia {\sl et al.},
Comput. Phys. Commun. {\bf 178}, 956-971 (2008).

\bibitem[\protect\citeauthoryear{Goodwin et al.}{1999}]{B5}
S.P.~Goodwin, P.~Thomas, A.J.~Barber, J.~Gribbin, L.I.~Onuora,
arXiv:astro-ph/9906187 (1999).

\bibitem[\protect\citeauthoryear{Perlmutter et al.}{1999}]{B6}
S.~Perlmutter {\sl et al.},
Astrophys. J. {\bf 517}, 565 (1999).

\bibitem[\protect\citeauthoryear{Szab\'o et al.}{2007}]{B9}
G.M.~Szab\'o, L.A.~Gergely and Z.~Keresztes,
PMC Physics A {\bf 1}, 8 (2007).

\bibitem[\protect\citeauthoryear{Komatsu et al.}{2009}]{B11}
E.~Komatsu {\sl et al.},
Astrophys. J. Suppl. Series {\bf 180}, 330-376 (2009). 

\bibitem[\protect\citeauthoryear{Serra \& Dominguez}{2011}]{C1}
A.L.~Serra and M.~Dominguez,
Mon. Not. R. Astron. Soc. {\bf 415}, L74-L77 (2011).

\bibitem[\protect\citeauthoryear{Dom\'\i{}nguez \& Ruiz}{2012}]{C2}
M.~Dom\'\i{}nguez, A.N.~Ruiz,
SOURCEAIP Conference Proceedings {\bf 1471}, 70 (2012).	

\bibitem[\protect\citeauthoryear{Bull et al.}{2012}]{Bulletal2012}
P.~Bull {\sl et al.}, 
Phys. Rev. {\bf D85}, 024002 (2012) 
\end{thebibliography}
